\documentclass[12pt,preprint]{aastex}
\usepackage{epsfig}

\def\gtrsim{\mathrel{\hbox{\rlap{\hbox{\lower4pt\hbox{$\sim$}}}\hbox{$>$}}}}
\def\lesssim{\mathrel{\hbox{\rlap{\hbox{\lower4pt\hbox{$\sim$}}}\hbox{$<$}}}}

\def\vkm{km s$^{-1}$}

\def\degree{$^\circ$}
\def\arcs#1{$#1''$}
\def\arcsa#1#2{$#1^{\prime\prime}_{^\textrm{.}}#2$}

\def\solarmass{$M_\odot$}

\def\mJyb{mJy beam$^{-1}$}

\def\tlabel#1{(\textit{#1})}

\def\cms{cm$^{-2}$}
\def\micron{$\mu$m}

\def\mHa{m_\textrm{\scriptsize H}}

\def\nHm{n_{\textrm{\scriptsize H}_2}}
\def\nHa{n_{\textrm{\scriptsize H}}}
\def\nHe{n_\textrm{\scriptsize He}}

\def\N2HP{N$_2$H$^+$}

\def\NH3{NH$_3$}

\def\putfig#1#2#3{\epsfig{scale=#1,angle=#2,figure=#3}}
\def\putfiga#1#2#3{}
\def\leftblank#1{}
\def\ssr{Space Science Reviews}

\newcounter{mfigure}[section]
\newenvironment{mfigure}[1][]{\refstepcounter{mfigure}\par\medskip
   \noindent \textbf{Figure~\themfigure. #1} \rmfamily}{\medskip}

\title{Stratified Distribution of Organic Molecules at the Planet-Formation
Scale in the HH 212 Disk Atmosphere}

\author{Chin-Fei Lee\altaffilmark{1,2}, Claudio Codella\altaffilmark{3,4}, Cecilia
Ceccarelli\altaffilmark{4}, and Ana L\'opez-Sepulcre\altaffilmark{4,5}}

\altaffiltext{1}{Academia Sinica Institute of Astronomy and Astrophysics,
P.O. Box 23-141, Taipei 106, Taiwan; cflee@asiaa.sinica.edu.tw}
\altaffiltext{2}{Graduate Institute of Astronomy and Astrophysics, National Taiwan
   University, No.  1, Sec.  4, Roosevelt Road, Taipei 10617, Taiwan}
\altaffiltext{3}{INAF, Osservatorio Astrofisico di Arcetri, Largo E. Fermi 5,
50125 Firenze, Italy}
\altaffiltext{4}{Univ. Grenoble Alpes, CNRS, Institut de
Plan\'etologie et d'Astrophysique de Grenoble (IPAG), 38000 Grenoble, France}
\altaffiltext{5}{
Institut de Radioastronomie Millim\'etrique (IRAM), 300 rue de la Piscine,
F-38400 Saint-Martin d'H\`eres, France}

\begin{document}

\def\NHtCHO{NH$_2$CHO}
\def\HtCO{H$_2$CO}
\def\DtCO{D$_2$CO}
\def\HttCO{H$_2$$^{13}$CO}
\def\CHtDOH{CH$_2$DOH}
\def\CHtOH{CH$_3$OH}
\def\tCHtOH{$^{13}$CH$_3$OH}
\def\CHtCHO{CH$_3$CHO}
\def\CHtDCN{CH$_2$DCN}
\def\uJyb{$\mu$Jy beam$^{-1}$}
\def\scnum#1#2{#1$\times10^{#2}$}

\begin{abstract}

Formamide (\NHtCHO{}) is considered an important prebiotic molecule because
of its potential to form peptide bonds.  It was recently detected in
the atmosphere of the HH 212 protostellar disk on the Solar-System scale
where planets will form.  Here we have mapped it and its potential parent
molecules HNCO and \HtCO{}, along with other molecules \CHtOH{} and
\CHtCHO{}, in the disk atmosphere, studying its formation mechanism. 
Interestingly, we find a stratified distribution of these molecules, with
the outer emission radius increasing from $\sim$ 24 au for \NHtCHO{} and
HNCO, to 36 au for \CHtCHO{}, to 40 au for \CHtOH{}, and then to 48 au for
\HtCO{}.  More importantly, we find that the increasing order of the outer
emission radius of \NHtCHO{}, \CHtOH{}, and \HtCO{} is consistent with the
decreasing order of their binding energies, supporting that they are
thermally desorbed from the ice mantle on dust grains.  We also find that
HNCO, which has much lower binding energy than \NHtCHO{}, has almost the
same spatial distribution, kinematics, and temperature as \NHtCHO{}, and is
thus more likely a daughter species of desorbed \NHtCHO{}.  On the other
hand, we find that \HtCO{} has a more extended spatial distribution with
different kinematics from \NHtCHO{}, thus questioning whether it can be the
gas-phase parent molecule of \NHtCHO{}.

\end{abstract}

\section{Introduction}

Formamide (\NHtCHO{}) is an interstellar complex organic molecule (iCOM,
referring to C-bearing species with six atoms or more)
\citep{Herbst2009,Ceccarelli2017} and a key precursor of more complex
organic molecules, that can lead to the origin of life,
because of its potential to form peptide bonds \citep{Saladino2012,Kahane2013,Lopez2019}.  It has been detected
in gas phase in hot corinos
\citep{Kahane2013,Coutens2016,Imai2016,Lopez2017,Bianchi2019,Hsu2022}, which
are the hot ($\gtrsim 100$ K) and compact ($\lesssim 100$ au) regions
immediately around low-mass (sun-like) protostars \citep{Ceccarelli2007}. 
The formamide origin is still under debate.  In principle, formamide could
be synthesized on the grain surfaces or in the gas phase.  Two routes have
been proposed in the first case: the hydrogenation of HNCO
\citep{Charnley2008} and the combination of the HCO and NH$_2$ radicals,
when they become mobile upon the warming of the dust by the protostar. 
However, the first route has been challenged by both experiments
\citep{Noble2015} and quantum chemical (QM) calculations \citep{Song2016}.
 Later, the hydrogenation of HNCO is found to be feasible and followed by H
abstraction of \NHtCHO{} in a dual-cycle consisting of H addition and H
abstraction \citep{Haupa2019}.  The second route (i.e., the combination
of HCO and NH$_2$) has also been challenged by QM calculations
\citep{Rimola2018} and found possible, even though it can also form
NH$_3$ $+$ CO in competition with the formamide \citep{Enrique-Romero2022}. 
In the gas-phase formation theory, it has been proposed that formamide is
formed by the gas-phase reaction of H$_2$CO with NH$_2$ \citep{Kahane2013}. 
This hypothesis was later challenged by \citet{Song2016}.  Nonetheless, QM
computations \citep{Vazart2016,Skouteris2017} coupled with astronomical
observations in shocked regions \citep{Codella2017} support this hypothesis. 
On the same vein, the observed deuterated isomers of formamide
(including NH$_2$CDO, cis- and trans-NHDCHO) \citep{Coutens2016} fits well
with the theoretical predictions of a gas-phase formation route
\citep{Skouteris2017}.  On the other hand, the observed high deuterium
fractionation of $\sim$ 2\% for the three different forms of formamide
(NH$_2$CDO, cis- and trans-NHDCHO) could also be consistent with the
formation in ice mantles on dust grains.

%On the other hand, the {\bf observed high D/H ratio
%of formamide} could also be consistent with the formation in ice mantles on
%dust grains.

The hot corino in the HH 212 protostellar system \citep{Codella2016} in
Orion at a distance of $\sim$ 400 pc is particularly interesting because
recent observations have spatially resolved it and found it to be an
atmosphere of a Solar-System scale protostellar disk around a protostar
\citep{Lee2017COM}.  This disk atmosphere is rich in iCOMs
\citep{Lee2017COM,Codella2018,Lee2019COM}, including formamide.  More
importantly, these iCOMs have a relative abundance similar to that of other
hot corinos \citep{Cazaux2003,Imai2016,Lopez2017,Bianchi2019,Manigand2020},
and even comets \citep{Biver2015}.   Therefore, the study of formamide
in protostellar disks is key to investigate the emergence of prebiotic
chemistry in nascent planetary bodies.   In this paper we
will study the origin and formation pathways of formamide in this
protostellar disk.

Previously, the HH 212 disk was mapped at a wavelength $\lambda \sim$ 0.85
mm \citep{Lee2017COM} covering one \NHtCHO{} line.  Here we map it at a
longer wavelength $\lambda \sim$ 1.33 mm, with spectral windows set up to
cover more \NHtCHO{} lines in order to derive the physical properties
of \NHtCHO{}.  This set up also covers the lines of HNCO and \HtCO{}, which
have not been reported before, in order to investigate the formation
pathways of \NHtCHO{}. At longer wavelength, since the continuum emission
of the disk is optically thinner, we can also map the molecular line
emission in the disk atmosphere closer to the midplane and the central
source.  Moreover, the deuterated species and $^{13}$C isotopologue of
\HtCO{} are also detected, allowing us to constrain the origin and correct
the optical depth of \HtCO{}, respectively.  In addition, \CHtOH{} and
\CHtCHO{} are also detected, allowing us to further constrain the formation
mechanism of \NHtCHO{}.  More importantly, with the recently updated binding
energies of these molecules, we can investigate the formation mechanism of
these molecules and the chemical relationship among them.

%\subsection{Distribution}

% Here Band6, 0.021x0.0162, PA=79.18, robust=-1.0
% In Lin et al. 2021, Band6 has 0.026x0.021, PA=60.3, robust=+0.5

\section{Observations} 

The HH 212 protostellar disk was observed with Atacama Large
Millimeter/submillimeter Array (ALMA) in Band 6 centered at a frequency of
$\sim$ 226 GHz (or $\lambda \sim$ 1.33 mm) in Cycle 5.  Project ID was
2017.1.00712.S.  Two observations were executed.  One was executed on 2017
October 04 in C43-9 configuration with 46 antennas for $\sim$ 18 mins on
source with a baseline length of 41.4 m to 15 km to achieve an angular
resolution of $\sim$ \arcsa{0}{02}.  The other was executed on 2017 December
31 in C43-6 configuration with 46 antennas for 9 mins on source with a
baseline length of 15.1 m to 2.5 km to recover a size scale up to $\sim$
\arcsa{1}{8}, which is 4 times the disk size.  The correlator was set up to
have 4 spectral windows (centered at 232.005, 234.005, 217.765, and 219.705
GHz) , each with a bandwidth of 1.875 GHz and 1920 channels, and thus a
spectral resolution of 0.976 MHz per channel, corresponding to $\sim$ 1.3
\vkm{} per channel.  The primary beam was $\sim$ \arcs{25}, much larger than
the disk size.

%b8 C43-9  Array, SB Spectral Setup
%(231.068-232.943 @ 232.005, 233.068-234.943 @ 234.005
%,216.827-218.702 @ 217.765 ,218.768-220.643 @ 219.705) 
% calc "(232.005+234.005+217.765+219.705)/4" ==  225.8700

%Sources
%0 	J0423-0120 	04:23:15.801 	-001.20.33.065 	ICRS 			1 	AMPLITUDE, ATMOSPHERE, BANDPASS, POINTING, WVR
%1 	J0541-0211 	05:41:21.696 	-002.11.08.387 	ICRS 			1 	PHASE, WVR
%2 	J0541-0541 	05:41:38.083 	-005.41.49.429 	ICRS 			1 	CHECK, WVR
%3 	HH_212 	05:43:51.411 	-001.02.53.167 	ICRS 			1 	ATMOSPHERE, TARGET

% file:///Users/cflee/ALMAHH212_Cy5/B6/b8/qa/pipeline-20171110T182654/html/stage1/t2-4m_details.html
% cont.map_Rob-1r

The data were calibrated with the CASA package version 5.1.1, with quasar J0423-0120 (a
flux of $\sim$ 0.93 Jy) as a passband and flux calibrator, and quasar
J0541-0211 (a flux of $\sim$ 0.096 Jy) as a gain calibrator.  Line-free
channels were combined to generate a visibility for the continuum centered
at 226 GHz.  We used a robust factor of $-$1.0 for the visibility weighting
to generate the continuum map with a synthesized beam of
\arcsa{0}{021}$\times$\arcsa{0}{016} at a position angle of $\sim$
79\degree{}.  The noise level is $\sim$ 20 \uJyb{} or 1.4 K.  The channel
maps of the molecular lines were generated after continuum subtraction. 
Using a visibility weighting of 0.5, the synthesized beam has a size of
\arcsa{0}{055}$\times$\arcsa{0}{042} at a position angle of $\sim$
49\degree{}.  The noise levels are $\sim$ 0.9 \mJyb{} (or $\sim$ 10 K) in
the channel maps.  The velocities in the channel maps are LSR velocities.

\section{Results}

The detected lines of \NHtCHO{} (16 lines), HNCO (7 lines), \HtCO{} (2
lines) as well as its doubly deuterated species \DtCO{} (2 lines) and
$^{13}$C isotopologue \HttCO{} (1 line),
\CHtOH{} (12 lines), and \CHtCHO{} (7 lines) are listed in Table
\ref{tab:lines}.  They have upper level energy $E_u \lesssim 500$ K, but
with $E_u < 120$ K for \CHtCHO{}, \HtCO{} as well as its deuterated species
and isotopologue.  In order to increase the sensitivity for better
detections, we divided them into 2 ranges of upper level energies: $E_u <
120$ K and $E_u > 120 K$, and then stacked them to produce the mean channel
maps, and then the total line intensity maps, and the position-velocity (PV)
diagrams.

\subsection{Stratified Distribution of Molecules}

%H2CO is organic
%HNCO is inorganic

Figure \ref{fig:contCOMs} shows the total line intensity maps (red contours)
of these molecules on top of the continuum map of the disk at $\lambda \sim$
1.33 mm, in order to pin point the location of these molecules in the disk and
the chemical relationship among them.  As shown in Figure
\ref{fig:contCOMs}c, the disk is nearly edge-on, with an equatorial dark
lane tracing the cooler midplane sandwiched by two brighter features
(outlined by the 4th and 5th contour levels)
on the top and bottom tracing the warmer surfaces, as seen before in continuum at a
shorter wavelength of $\sim$ 0.85 mm \citep{Lee2017Disk}.

As can be seen, the emission structure of a given molecule is similar in
different $E_u$ ranges, suggesting that it is more dominated by the
distribution of the molecule than the upper energy level.  After stacking
the lines, we achieved a better sensitivity in \NHtCHO{} than that in the
previous observations obtained at higher resolution \citep{Lee2017COM}, and
detected \NHtCHO{} not only in the lower disk atmosphere, but also in the
upper disk atmosphere.  More importantly, we can better pinpoint its
emission and found it to be in the inner disk where the disk is warmer. It is
brighter in the lower disk atmosphere, with two emission peaks clearly seen
in the map with $E_u > 120$ K (see Figure \ref{fig:contCOMs}b).  HNCO is
detected with the spatial distribution and radial extent consistent with
\NHtCHO{}.  Looking back at the previous results of other iCOMs detected at
higher frequency of $\sim$ 346 GHz \citep{Lee2019COM}, we find that t-HCOOH
was also detected with the spatial distribution and radial extent consistent
with \NHtCHO{} \cite[see Figure \ref{fig:contCOMs}f adopted
from][]{Lee2019COM}.  Notice that the radial distribution of molecular gas
detected at higher frequency can also be compared with that here at lower
frequency, because the optical depth of the underlying continuum of the
dusty disk mainly affects the vertical distribution (i.e., height) of
molecular gas in the atmosphere (see Section \ref{sec:midplane}).  On the
other hand, \HtCO{} is only detected with $E_u < 70$ K, and its emission
extends further out in radial direction beyond the centrifugal barrier (CB)
(Figure \ref{fig:contCOMs}g).  The emission is also detected at a larger
distance from the disk miplane and extends away from the disk atmosphere,
overlapping with the base of the SO disk wind \citep{Lee2021DW} and thus
tracing the wind from the disk.  Its deuterated species \DtCO{} is detected
mainly in the disk atmosphere, also at a larger distance from the midplane
and a larger radius from the central protostar than \NHtCHO{}.  On the other
hand, the emission of the $^{13}$C isotopologue \HttCO{} is very faint and
mainly detected in the disk atmospheres.  \CHtOH{} is detected in the
atmosphere extending out to the CB, as found before
\citep{Lee2017COM,Lee2019COM}.  The emission also extends away from the disk
midplane, suggesting that part of it also traces the wind from the disk.  As
for \CHtCHO{}, the emission is mainly detected in the disk atmosphere and
extends radially toward the CB.  In summary, \NHtCHO{}, HNCO, \DtCO{},
\HttCO{} and \CHtCHO{} trace mainly the disk atmosphere, while \HtCO{} and
\CHtOH{} trace not only the disk atmosphere, but also the disk wind.

%, at which the optical depth of the underlying continuum is higher,

%0.036x 0.3 vs 0.055x0.042

We can also measure the vertical height of these molecules (using lines with
$E_u < 120$ K) along the jet axis in the lower atmosphere where the emission
is brighter, and find it to be $\sim$ 15, 19, 20, 24, and 26 au,
respectively for \NHtCHO, HNCO, \CHtCHO{}, \CHtOH{}, and \HtCO{}.  We will
discuss the vertical height later with the outer radius of these molecules
measured from the PV diagrams.

\subsection{Kinematics}

%At $Eu<250$ K, 2 HNCO lines are only seperated by 4.5 \vkm{}, can not be separated in PV.

The spatio-kinematic relationship among these molecules can be studied with
the PV diagrams cut across the upper and lower disk atmospheres, as shown in
Figure \ref{fig:pv_atms}.  Here we use the emission with $E_u < 120$ K,
where all molecules are detected.  In addition, this emission is expected to
trace the lowest temperature and thus the outermost radius at which the
molecules start to appear.  Previously, the disk was found to be rotating
roughly with a Keplerian rotation due a central mass of $\sim$ 0.25
\solarmass{} (including the masses of the central protostar and the
disk) \citep{Codella2014,Lee2017COM}.  Therefore, the associated Keplerian
rotation curves (blue curves) are plotted here for comparison.

The emissions of these molecules trace the disk atmosphere within the CB and
are thus enclosed by the Keplerian rotation curves.  In the upper disk
atmosphere, their emissions form roughly linear PV structures (as marked by
the magenta lines), indicating that they arise from rings rotating at
certain velocities.  For edge-on rotating rings, the radial velocity
observed along the line of sight is proportional to the position offset from
the center, forming the linear PV structures. Interestingly, the PV
structures of HNCO and t-HCOOH are aligned with those of \NHtCHO{}, and the
PV structures of \DtCO{} are roughly aligned with those of \CHtOH{}.  Except
for these similarities, different molecules have different velocity
gradients connecting to different locations of the Keplerian curves,
indicating that they arise from rings at different disk radii.  From the
location of their PV structure on the Keplerian curve, we find that the disk
radius of these molecules increases from $\sim$ 24 au for
\NHtCHO{}/HNCO/t-HCOOH, to $\sim$ 36 au for \CHtCHO{}, to $\sim$ 40 au for
\CHtOH{}/\DtCO{}, and then to $\sim$ 48 au for \HtCO{}.  This trend is the
same as the increasing order of the vertical height measured earlier for
these molecules, indicating that the height increases with increasing
radius, as expected for a flared disk in hydrostatic equilibrium.  Plotting
the velocity gradients in the upper disk atmosphere onto the lower disk
atmosphere, we find that the emission detected in the upper disk atmosphere
is actually only from the outer radius where their emission start to appear,
and the emission also extends radially inward to where \NHtCHO{} is
detected.  Since the nearside of the disk is tilted slightly downward to the
south, the emission in the upper disk atmosphere further in is lost due to
the absorption against the bright and optically thick continuum emission of
the disk surface (see Figure 9b in Lee et al.  2019).  Note that for
\NHtCHO{}/HNCO/t-HCOOH, there seems to be a small velocity shift of $\sim$
0.5 \vkm{} between the upper and lower disk atmosphere.  This velocity shift
could suggest an infall (or accretion) velocity of $\sim$ 0.25 \vkm{}, which
is $\sim$ 8\% of the rotation velocity at $\sim$ 24 au.  However,
observations at higher spectral and spatial resolution are needed to verify
this possibility.

\subsection{Physical Properties in the Disk Atmosphere}

%(with a size of \arcsa{0}{17}$\times$\arcsa{0}{05} or 68 au $\times$ 20 au)

In order to understand the nature and spatial origin of the detected
methanol, we analyzed the observed methanol lines (Table \ref{tab:lines})
via a non-LTE Large Velocity Gradient (LVG) approach, using
the code \textsc{grelvg}, initially developed by \citet{Ceccarelli2003}.  We
used the collisional coefficients of methanol with para-H$_2$, computed by
\citet{Rabli2010} between 10 and 200 K for the J$\leq$15 levels and provided
by the BASECOL database \citep{Dubernet2012,Dubernet2013}.  We assumed an
A-/E- CH$_3$OH ratio equal to 1.  To compute the line escape probability as
a function of the line optical depth we adopted the semi-infinite slab
geometry \citep{Scoville1974} and a linewidth equal to 4 km~s$^{-1}$,
following the observations.

We ran several grids of models to sample the $\chi^2$ surface in the
parameter space.  Specifically, we varied the methanol column density
N(A-CH$_3$OH) and N(E-CH$_3$OH) simultaneously from $2\times 10^{15}$ to
$1\times 10^{19}$ cm$^{-2}$ (with a step of a factor of 2), the H$_2$
density $\nHm$ from $10^{6}$ to $10^{9}$ cm$^{-3}$ (with a step of a
factor of 2) and the gas temperature T from 50 to 120 K (with a step
of 5 K).  We then fit the measured the velocity-integrated line intensities
($W=\int T_B dv$ with $T_B$ being the brightness temperature) by comparing
them with those predicted by the model, leaving N(A-CH$_3$OH) and
N(E-CH$_3$OH), $\nHm$, and $T$ as free parameters.  Given the limitation on
the J level ($\leq$15), we used only seven of the twelve detected
methanol lines with $E_u < 200$ K for the LVG fitting. We considered the
line intensities at the emission peak (marked by a blue circle in Figure
\ref{fig:contCOMs}j) in the lower disk atmosphere, as listed in Table
\ref{tab:lines} .

The results of the fit are shown in Figure \ref{fig:LVG}.  The best fit
gives the following values, where the errors are estimated considering the
1$\sigma$ confidence level and the uncertainties of $\sim$ 40\% in our
measurements: N(CH$_3$OH)$=$N(A-CH$_3$OH)$+$N(E-CH$_3$OH)$\sim
1.6^{+4.4}_{-0.8} \times 10^{18}$ cm$^{-2}$; $\nHm \sim 10^{9}$ cm$^{-3}$,
which should be the lower limit because it is in the LTE regime at this
density; and $T \sim 75\pm20$ K.  The lines are predicted to be all
optically thick with the lowest line opacity $\tau \sim 1$ for the line at
234.699 GHz and the highest $\tau\sim 19$ for the line at 218.440 GHz, and
$\tau$=3--10 for the other lines.  We also derived the excitation
temperature and column density from rotation diagram using the remaining
five transition lines with $E_u > 200 $K (see Figure \ref{fig:LVG}c),
assuming optically thin emission and LTE \citep{Goldsmith1999}.  In
particular, we fit the data with a linear equation, and then derived
the temperature from the negative reciprocal of the slope and the column density
from the y-intercept. We found that $T\sim 109\pm31$ K and
N(CH$_3$OH)$=(1.4\pm0.7)\times 10^{18}$ cm$^{-2}$.  Taking the mean values
from the two methods, we have $T\sim 92^{+48}_{-37}$ K and
N(CH$_3$OH)$=1.5^{+4.5}_{-0.8}\times 10^{18}$ cm$^{-2}$.  Notice that
previous LTE estimation of excitation temperature of \CHtOH{} and \CHtDOH{}
together from rotation diagram was pretty uncertain, with a value of
165$\pm$85 K \citep{Lee2017COM}, due to a large scatter of the data points. 
More importantly, the excitation temperature was also overestimated because
almost all the lines had E$_u$ $<$ 200 K and were thus likely optically
thick.

For less abundant molecules detected with a broad range of $E_u$, such as
\NHtCHO{} and HNCO, the mean excitation temperature and column density of
the molecular lines in the disk atmosphere can be roughly estimated from
rotation diagram assuming optically thin emission and LTE
\citep{Goldsmith1999}.  We used the brighter emission in the lower disk
atmosphere.  Table \ref{tab:lines} lists the integrated line intensities
averaged over a rectangular region (with a size of 68 au $\times$ 20 au)
that covers most of the emission in the lower atmosphere, measured with a
cutoff of 2$\sigma$.  Figure \ref{fig:popdia} shows the resulting rotation
diagrams for \NHtCHO{} and HNCO.  The blended lines of \NHtCHO{} are
excluded from the diagram.  The HNCO line at the lowest $E_u$ (marked with
an open square) seems to be optically thick with an intensity much lower
than the line next down the $E_u$ axis, and is thus excluded from the
fitting.  For \NHtCHO{} and HNCO, we fit the data points to obtain the
temperature and column density.  It is interesting to note that \NHtCHO{}
and HNCO have roughly the same excitation temperature of $\sim$ $226\pm130$
K, although with a large uncertainty.  On the other hand, since \HtCO{},
\DtCO{}, and \CHtCHO{} are only detected with a narrow range of $E_u < 120$
K and their emission can be optically thick there, we can not derive their
excitation temperature from the rotation diagram.   In addition, \HtCO{}
and \DtCO{} are only detected with two lines. Also, \HttCO{} is only
detected with one line.  Since \DtCO{} has roughly the same radial extent as
\CHtOH{}, it is assumed to have an excitation temperature of 92  K, the same
as that found for \CHtOH{}.  Since \HtCO{} has a slightly larger radius
than \CHtOH{}, it and its $^{13}$C isotopologue \HttCO{} are assumed to have
an excitation of 60 K. \CHtCHO{} has a smaller radial extent than \CHtOH{}
and is thus assumed to have an excitation temperature of 100 K.

The resulting excitation temperature and column density are listed in Table
\ref{tab:colabun}.  In addition, the abundance of these molecules are also
estimated by dividing the column density of the molecules by the mean H$_2$
column density derived from a dusty disk model \citep{Lee2021Pdisk} in the
same region, which is found to be $\sim$ \scnum{1.08}{25} \cms{}.  This disk
model was constructed before to reproduce the continuum emission of the disk
at $\lambda \sim$ 850 \micron{} \citep{Lee2021Pdisk} and it can also roughly
reproduce the continuum emission of the disk observed here at $\lambda \sim$
1.33 mm \citep{Lin2021}.  Since \HtCO{} and \DtCO{} lines are each detected
with two lines that are likely optically thick, their lines at higher E$_u$
are used to derive the lower limit of their column density.  Indeed, the
\HtCO{} column density can be better derived from the \HttCO{} line assuming
[$^{12}$C]/[$^{13}$C] ratio of $\sim$ 50, as estimated in the Orion Complex
\citep{Kahane2018}.  As can be seen from Table \ref{tab:colabun}, the
\HtCO{} column density derived this way is $\sim$ 3 times that derived from
the \HtCO{} lines.  Thus, the deuteration of \HtCO{}, i.e., the abundance
ratio [\DtCO{}]/[\HtCO{}], is $\gtrsim$ 0.053.  As for \CHtCHO{}, we
fixed its temperature to 100 K by fixing the negative reciprocal of the
slope in the linear equation and then derived its column density from the
y-intercept of the linear fit to the rotation diagram, as shown in Figure
\ref{fig:popdia}c.

%Here we derive the mean excitation temperature and column density in the
%disk atmosphere by fitting the rotation diagram of the molecular lines. 
%This diagram plots the column density per statistical weight in the upper
%energy state in the optically thin limit, $N_u^\textrm{\scriptsize
%thin}/g_u$, versus the upper energy level $E_u$ of the lines.  Here
%$N_u^\textrm{\scriptsize thin}=(8\pi k\nu^2/hc^3 A_{ul}) W$, where the
%integrated line intensity $W =\int T_B dv$ with $T_B$ being the brightness
%temperature.

\section{Discussion}

\subsection{Lack of Molecular Emission in Disk Midplane} \label{sec:midplane}

As discussed in \citet{Lee2017COM,Lee2019COM}, the lack of molecular
emission in the disk midplane can be due to an exponential attenuation by
the high optical depth of dust continuum.  Figure \ref{fig:conttau}a shows
the optical depth of the dust continuum at 1.33 mm derived from the dusty
disk model that reproduced the thermal emission of the disk
\citep{Lee2021Pdisk}.  As can be seen, the optical depth is $\gtrsim$ 3
toward the midplane within the CB, where no molecular emission is detected,
supporting this possibility.  The faint \NHtCHO{} and \CHtOH{} emission
detected in the midplane likely comes from the upper and lower disk
atmospheres due to the beam convolution.  However, the \HtCO{} emission in
the midplane near the CB (see Figure \ref{fig:contCOMs}g) should be real
detection because the optical depth of the dust continuum decreases 
to smaller than 3 at the edge.

\subsection{Distribution of Molecules and Binding Energy}

As discussed earlier, a stratification is seen in the distribution of
molecules in the disk atmosphere, with the outer disk radius decreasing from
\HtCO{}, to \CHtOH{}, to \CHtCHO{}, and then to \NHtCHO{}/HNCO/HCOOH, as
shown in Figure \ref{fig:conttau}b together with the temperature structure
of the dusty disk model \citep{Lee2021Pdisk}.  Similar stratification of
\HtCO{}, \CHtCHO{}, and \NHtCHO{} has been seen toward the bow shock region
B1 in the young protostellar system L1157 \citep{Codella2017}.  That shock
region is divided into 3 shock subregions, with shock 1 in the bow wing,
shock 3 in the bow tip, and shock 2 in between.  The authors interpreted
that shock 1 is the youngest shock while shock 3 is the oldest.  They found
that the observed decrease in abundance ratio [\CHtCHO{}]/[\NHtCHO{}] from
shock 3 to shock 2 and to shock 1 can be modeled if both \NHtCHO{} and
\CHtCHO{} are formed in gas phase.

%{\bf (From their Table 3, I derived a much higher value of 6509 K for HNCO
%(see Table \ref{tab:BE}).)}

% supporting that these molecules could have already formed in ice mantle on
%dust grains and are now thermally desorbed into gas phase.

Here in the HH 212 disk, since the temperature of the atmosphere is expected
to increase inward toward the center as the disk (Figure
\ref{fig:conttau}b), the stratification in the distribution of these
molecules could be related to their binding energy (BE) (and thus
sublimation temperature).  Table \ref{tab:BE} lists the recently computed BE
for these molecules.  For consistent comparison, we adopt the values
obtained from similar methods on amorphous solid water ice
\citep{Ferrero2020,Ferrero2022}.  Since HNCO was not included in those
studies, we adopt its value from \citet{Song2016}.  Notice that different
methods can result in different BE, e.g., HCOOH was found to have a BE value
of less than 5000 K on pure ice \citep{Kruckiewicz2021}, significantly lower
than that adopted here. As can be seen, the increasing order of the
observed outer radius of \NHtCHO{}/t-HCOOH, \CHtOH{}, and \HtCO{} is
consistent with the decreasing order of their BE, indicating that these
molecules are thermally desorbed from the ice mantle on dust grains.  Notice
that this does not necessarily mean that these molecules are formed in the
ice mantle, because the density in the disk is so high that even if the
molecules are formed in the gas phase they freeze-out quickly and are,
therefore, only detected in regions where the dust temperature is larger
than the sublimation temperature.  As for HNCO and \CHtCHO{}, their outer
radii do not fit in to those of \HtCO{}, \CHtOH{}, and \NHtCHO{} based on
their BE and they can form in gas phase from other species.   On the
other hand, HNCO and HCOOH may come from the decomposition of the desorbed
organic salts (NH$_4^+$OCN$^-$ and NH$_4^+$HCOO$^-$), which have similar BE to
that of \NHtCHO{} \citep{Kruckiewicz2021,Ligterink2018}.  Further work is
needed to check this possibility.

Previously at $\sim$ \arcsa{0}{15} (60 au) resolution,
\citet{Codella2018} detected deuterated water around the disk.  Although the
deuterated water was found to have an outer radius of $\sim$ 60 au, its
kinematics was found to be consistent with that of the centrifugal barrier
at $\sim$ 44 au.  More importantly, since water has a BE similar to that of
\HtCO{} (see Table \ref{tab:BE}), it is likely that water, like \HtCO{}, is
also desorbed from the ice mantle on the dust grains.  Thus the water
snowline can be located around or slightly outside the centrifugal barrier.

%Higher resolution observations are needed to better resolve the outer radius of
%the deuterated water in order to confirm it.}

%Additional supporting evidence for \NHtCHO{} to have formed in ice mantle is
%that t-HCOOH, which has roughly the same BE as that of \NHtCHO{}, shows a
%similar distribution and kinematics to \NHtCHO{}.  Thus, both t-HCOOH and
%\NHtCHO{} could have formed in ice mantle and then are thermally desorbed at
%the same radius.  As for HNCO and \CHtCHO{}, their outer radii do not fit in
%to those of \HtCO{}, \CHtOH{}, and \NHtCHO{} based on their BE and they can
%form in gas phase from other species.

%{\bf This could be
%underestimated, only a fraction of \CHtCHO{} is desorbed at $\sim$ 100 K
%while 40\% of the molecules are retained by fluffy grains of the order of
%100 μm up to temperatures of 190$-$210 K \citep{Corazzi2021}.}

% {\bf Notice that however, the binding energy of
%\CHtCHO{} could be underestimated.  In Table \ref{tab:BE}, we also calculate
%the corresponding sublimation temperature assuming an age of 10$^6$ yrs, as
%if the molecules were formed in the prestellar core and then brought in to
%the disk, as discussed below.  Previous study of this molecule found that
%only a fraction of \CHtCHO{} was desorbed at 100 K \citep{Corazzi2021} and
%40\% of it is retained on fluffy grains of the order of 100 micron up to
%temperatures of 190$-$210 K.  } Since these temperatures
%are larger than the estimated kinetic temperature (80 K) of \CHtOH{},
%it is thus reasonable for \CHtCHO{} to have an outer radius smaller than \CHtOH{}.

\subsection{Centrifugal Barrier and \HtCO{} and \CHtOH{}}

The high deuteration of \HtCO{} (with [\DtCO{}]/[\HtCO{}] $\gtrsim$ 0.053)
and methanol (with [\CHtDOH]/[\CHtOH] $\sim$ 0.12) \citep{Lee2019COM}
supports that both are originally formed in ice.  These ratios of
[\DtCO{}]/[\HtCO{}] and [\CHtDOH]/[\CHtOH] are consistent with those found
in prestellar cores to Class I sources \cite[references
therein]{Mercimek2022}.  It is possible that \HtCO{} is formed by
hydrogenation to CO frozen in the ice mantle on dust grains and then
\CHtOH{} is formed from it with the addition of two H atoms
\citep{Charnley2004}.  The derived kinetic temperature of \CHtOH{} agrees
with the sublimation temperature, also supporting that the methanol is
thermally desorbed into gas phase.  \HtCO{} and \CHtOH{} are detected with
the outer radius near the CB where an accretion shock is expected as the
envelope material flows onto the disk \citep{Lee2017COM}, suggesting that
they are desorbed into gas phase due to the heat produced by the shock
interaction.  It is possible that they were already formed in the ice mantle
on dust grains in the collapsing envelope stage and then brought in to the
disk \citep{Herbst2009,Caselli2012}.  \HtCO{} has a lower sublimation
temperature than \CHtOH{}, and thus can be desorbed into gas phase further
out beyond the CB.  Interestingly, both \HtCO{} and \CHtOH{} also extend
vertically away from the disk surface, and thus can also trace the disk wind
as SO \citep{Tabone2017,Lee2018DW,Lee2021DW}.  In addition, since
\HtCO{} has an outer radius outside the centrifugal barrier, it may also
trace the wind from the innermost envelope transitioning to the disk,
carrying away angular momentum from there.

% Instead, \HtCO{} in the disk atmosphere shows similar distribution and
%kinematics to \CHtOH{}, suggesting a chemical link between the two.

%  The excitation temperature of methanol also agrees with the
%desorption temperature (128 to 162 K, with a peak at 150
%K)\citep{Lopez2019}, indicating that the methanol is desorbed into gas phase
%around the centrifugal barrier where an accretion shock is formed as the
%envelope material flows to the disk\cite{Lee2017COM}.  Interestingly, this
%derived excitation temperature is consistent with the surface temperature
%(145 K) of the disk model\citep{Lee2021Pdisk} and the expected postshock
%temperature ($\sim$ 135 K) at the centrifugal barrier (see Methods).  The
%excitation temperature of \HtCO{} can not be derived due to insufficient
%lines.  The thermal desorption temperature of \HtCO{} is $\sim$ 110 K from
%icy grain\citep{Noble2012}, thus \HtCO{} can be desorbed into gas phase near
%the centrifugal barrier.  It also extends inward to inner disk in the disk
%atmosphere due to desorption by accretion heating, stellar radiation
%heating, and disk-wind interaction\citep{Tabone2017,Lee2018DW,Lee2021DW}. 
%Note that \HtCO{} is also detected near the centrifugal barrier near the
%midplane where the temperature is lower than the surface, supporting that
%\HtCO{} can be desorbed at lower temperature than methanol.

\subsection{Formamide, HNCO, and \HtCO{}}

%Although HNCO has been detected in cold dark cloud \citep{Marcelino2009}, it
%is not detected here in the outer part of the disk where the temperature is
%low. 

HNCO not only has similar spatial distribution and kinematics, but also has
a similar excitation temperature to \NHtCHO{}, though with a large
uncertainty.  In addition, the abundance ratio of HNCO to \NHtCHO{} agrees
well with the nearly linear abundance correlation found before across
several orders of magnitude in molecular abundance \citep{Lopez2019}.  All
these suggest a chemical link between the two molecules.  However, as
discussed earlier based on the BE sequence, HNCO itself is likely formed in
gas phase but not desorbed from ice mantle, unless the BE of HNCO is
significantly underestimated.  In particular, although HNCO has a much lower
BE than \NHtCHO{}, it is detected only in the inner and warmer disk where
\NHtCHO{} is detected, but not detected in the outer part of the disk where
the temperature is lower.  Thus, our result implies that HNCO, instead of
being parent molecule, is likely a daughter molecule of \NHtCHO{} and formed
in gas phase.  One possible reaction is \NHtCHO{} $+$ H $\rightarrow$ HNCO
\citep{Haupa2019}. It is also possible that HNCO
is formed by destructive gas-phase ion-molecule interactions with
amides (also larger amides than \NHtCHO{}) \citep{Garrod2008,Tideswell2010}.

It has also been proposed that formamide can be formed from formaldehyde
(\HtCO) in warm gas through the reaction \HtCO{} $+$ NH$_2$ $\rightarrow$
\NHtCHO{} $+$ H \citep{Kahane2013,Vazart2016,Codella2017,Skouteris2017}. 
However, we find that \HtCO{} has a more extended distribution with
different kinematics from formamide, and is thus unclear if it can be the
parent molecule in gas phase.  Unfortunately, we have no information on the
other reactant, NH$_2$.  Very likely it is the product of sublimated NH$_3$
\citep{Codella2017}, whose binding energy (see Table \ref{tab:BE}) is
larger than that of \HtCO{}, which may explain why formamide is not present
where \HtCO{} is.  
 In conclusion, based on the current observations, it is not possible to
constrain the formation route of formamide in the disk atmosphere of HH 212. 
Nonetheless, our work has added precious information about the formation
route of formamide in disk atmosphere, complementing those in different
environments, e.g., the L1157 shock \citep{Codella2017}.

\acknowledgements We thank the anonymous reviewers for their insightful
comments. This paper makes use of the following ALMA data:
ADS/JAO.ALMA\#2017.1.00712.S.  ALMA is a partnership of ESO (representing
its member states), NSF (USA) and NINS (Japan), together with NRC (Canada),
NSC and ASIAA (Taiwan), and KASI (Republic of Korea), in cooperation with
the Republic of Chile.  The Joint ALMA Observatory is operated by ESO,
AUI/NRAO and NAOJ.  C.-F.L.  acknowledges grants from the Ministry of
Science and Technology of Taiwan (MoST 107-2119-M-001-040-MY3,
110-2112-M-001-021-MY3) and the Academia Sinica (Investigator Award
AS-IA-108-M01).  CC acknowledges the funding from the European Union’s
Horizon 2020 research and innovation programs under projects
“Astro-Chemistry Origins” (ACO), Grant No 811312; the PRIN-INAF 2016 The
Cradle of Life - GENESIS-SKA (General Conditions in Early Planetary Systems
for the rise of life with SKA); the PRIN-MUR 2020 BEYOND-2P (Astrochemistry
beyond the second period elements), Prot.  2020AFB3FX.

%\clearpage
%\newcommand\aap{{A\&A}}%M
%\newcommand\apjl{{ApJL}}%M
%\newcommand\apj{{ApJ}}%M
%\newcommand\apjs{{ApJS}}%M
%\newcommand\aj{{AJ}}%M
%\newcommand\araa{{ARAA}}%M
%\newcommand\nat{{Nature}}%M
%\newcommand\mnras{{MNRAS}}%M
%\newcommand\aapr{{A\&A Rev}}%M

%\begin{addendum}

%\item[Author Contributions]

%All other coauthors contribute to scientific discussion.

%\end{addendum}

%\clearpage

\def\tlabel#1{{\bf #1}}

\begin{figure} %[!hbp]
\centering
\putfig{1.1}{270}{f1.eps} %linemaps_mE.eps}
\figcaption[]
{Total line intensity maps of \NHtCHO{}, HNCO, t-HCOOH, \HtCO{}, \DtCO{}, \HttCO{},
\CHtOH{}, and \CHtCHO{} (red contours) on top of the 1.33 mm continuum map 
(gray image with black contours) toward the HH 212 disk.
The maps are rotated clockwise by $\sim$ 23\degree{} so that the disk is aligned
horizontally for easy view.  
For the continuum, the contours start from 10 K with a step of 15 K.
The line intensity maps are integrated over
velocity within $\sim$ 4.0 \vkm{} of the systemic velocity (which is $\sim
1.7\pm0.1$ \vkm{} LSR \citep{Lee2014}).  As described in the text,
we divided the lines into 2 ranges of upper level energies:
$E_u < 120$ K and $E_u > 120$ K, and then stacked them for each species for higher 
sensitivity. The asterisk marks the possible position of the central
protostar.  The vertical cyan lines mark the centrifugal barrier. The contours start at 3 $\sigma$
with a step of 2 $\sigma$, where $\sigma$ are (a) 1.5, (b) 0.8, (d) 2.1, (e) 1.3, (f) 4.0,
(g) 2.4, (h) 2.6, (i) 3.4, (j) 1.6, (k) 1.8, and (l) 1.4 \mJyb{}.
The blue circle in (j) marks the emission peak used to derive the physical properties of \CHtOH{}.
\label{fig:contCOMs}}
\end{figure}

\begin{figure} %[!hbp]
\centering
\putfig{0.75}{0}{f2.eps} %pv_ApJL.eps}
\figcaption[]
{Position-velocity diagrams of various molecular lines
cut across the upper and lower atmospheres of the disk.
The contours start from 3 $\sigma$ with a step of 2
$\sigma$, where $\sigma$ are 0.36, 0.51, 0.70, 0.51, 0.43, 0.70, and 0.22 \mJyb{} for
\NHtCHO{}, HNCO, t-HCOOH, \HtCO{}, \CHtOH{}, \DtCO{}, and \CHtCHO{}, respectively.
The vertical lines mark the systemic velocity.
The horizontal lines mark the center of the atmosphere.
We also plot the Keplerian
rotation curves (blue curves) due to a central mass of 0.25
\solarmass{} \citep{Lee2017COM}. The magenta lines roughly mark the linear velocity
gradients of the PV structures in the upper atmosphere.
\label{fig:pv_atms}}
\end{figure}

\begin{figure}
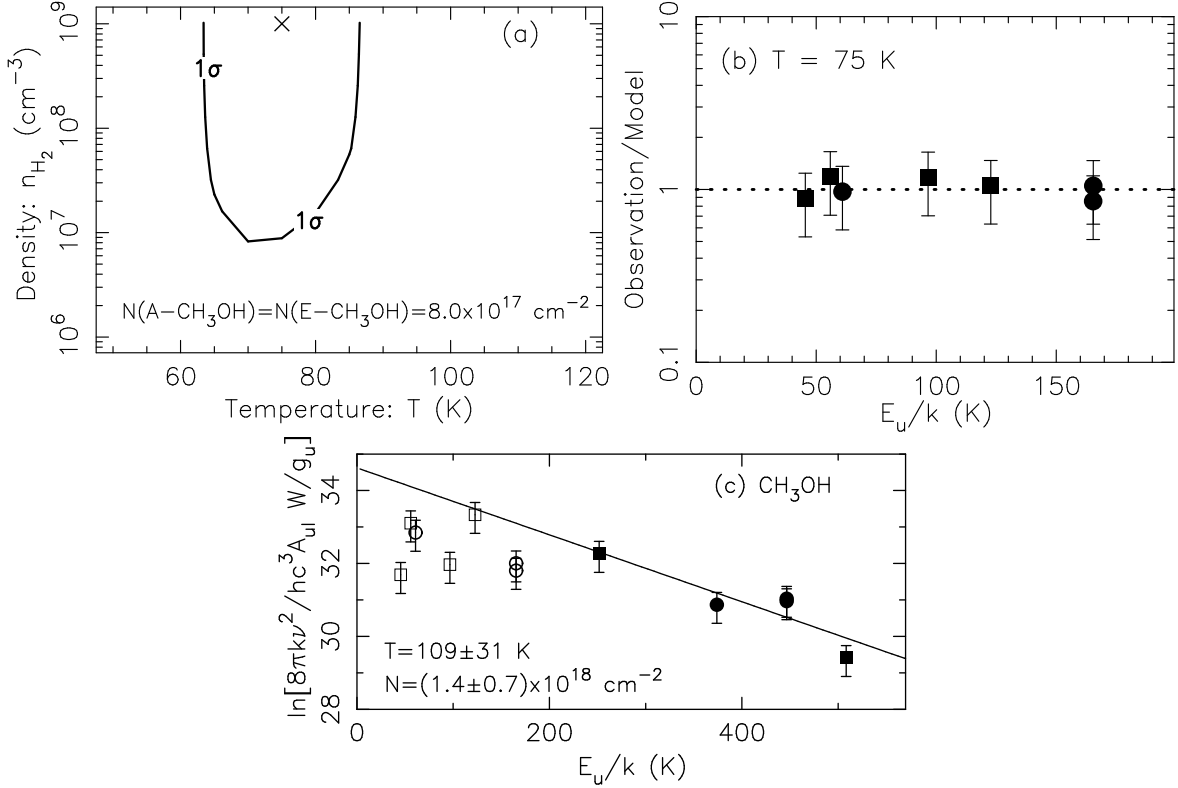
 %[!hbp]
\centering
\putfig{0.6}{270}{f3a.eps} % chi2.eps
\putfig{0.6}{270}{f3b.eps} % CH3OH_model_log.eps
\putfig{0.7}{270}{f3c.eps} % population_diagram2.eps
\figcaption[]
{Physical properties of \CHtOH{} toward the disk atmosphere. (a) and (b) show the
results of the non-LTE LVG analysis using the lines with $E_u < 200$ K
with the {\sc grelvg} code \citep{Ceccarelli2003}, while (c) shows the
result of rotation diagram using the lines with $E_u > 200$ K.  The line intensity is
extracted from an emission peak in the lower disk atmosphere, as marked
by a blue circle in Figure \ref{fig:contCOMs}j.
(a) Density-Temperature $\chi^2$ contour plot.  The contour
represents 1$\sigma$ confidence level, assuming the best fit values of
N(A-CH$_3$OH) and N(E-CH$_3$OH).  The best fit solution is marked by the
cross.  (b) Ratio between the observed line intensities
(circles for A-type and squares for E-type) with those of the best fit as a
function of E$_u$. (c) Rotation diagram of \CHtOH{} with the same symbols.
The solid line is a linear fit to the data. 
The optically thick lines 
are marked with open squares and circles and excluded from the fitting.
\label{fig:LVG}}
\end{figure}

\begin{figure}
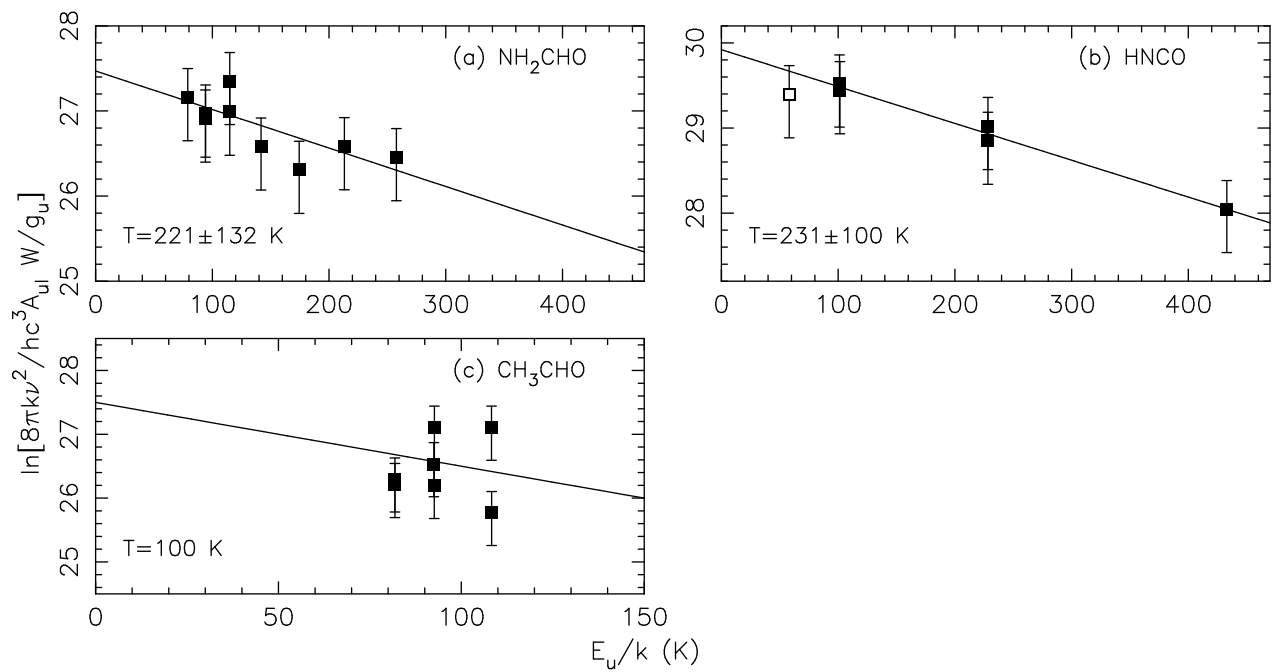
 %[!hbp]
\centering
\putfig{0.7}{270}{f4.eps} % population_diagram.eps}
\figcaption[]
{Rotation diagrams for molecular lines of \NHtCHO{}, HNCO, and \CHtCHO{}.
The diagrams are derived from the line intensities in the lower disk atmosphere listed
in Table \ref{tab:lines}.
The error bars show the uncertainty in our measurements, which are assumed 
to be 40\% of the data values.
The solid line is a linear fit to the data. The blended lines of
\NHtCHO{} are excluded from the diagram.  The optically thick line of HNCO
are marked with an open square and excluded from the fitting. 
For \CHtCHO{}, the temperature is fixed at 100 K.
\label{fig:popdia}}
\end{figure}

\begin{figure}
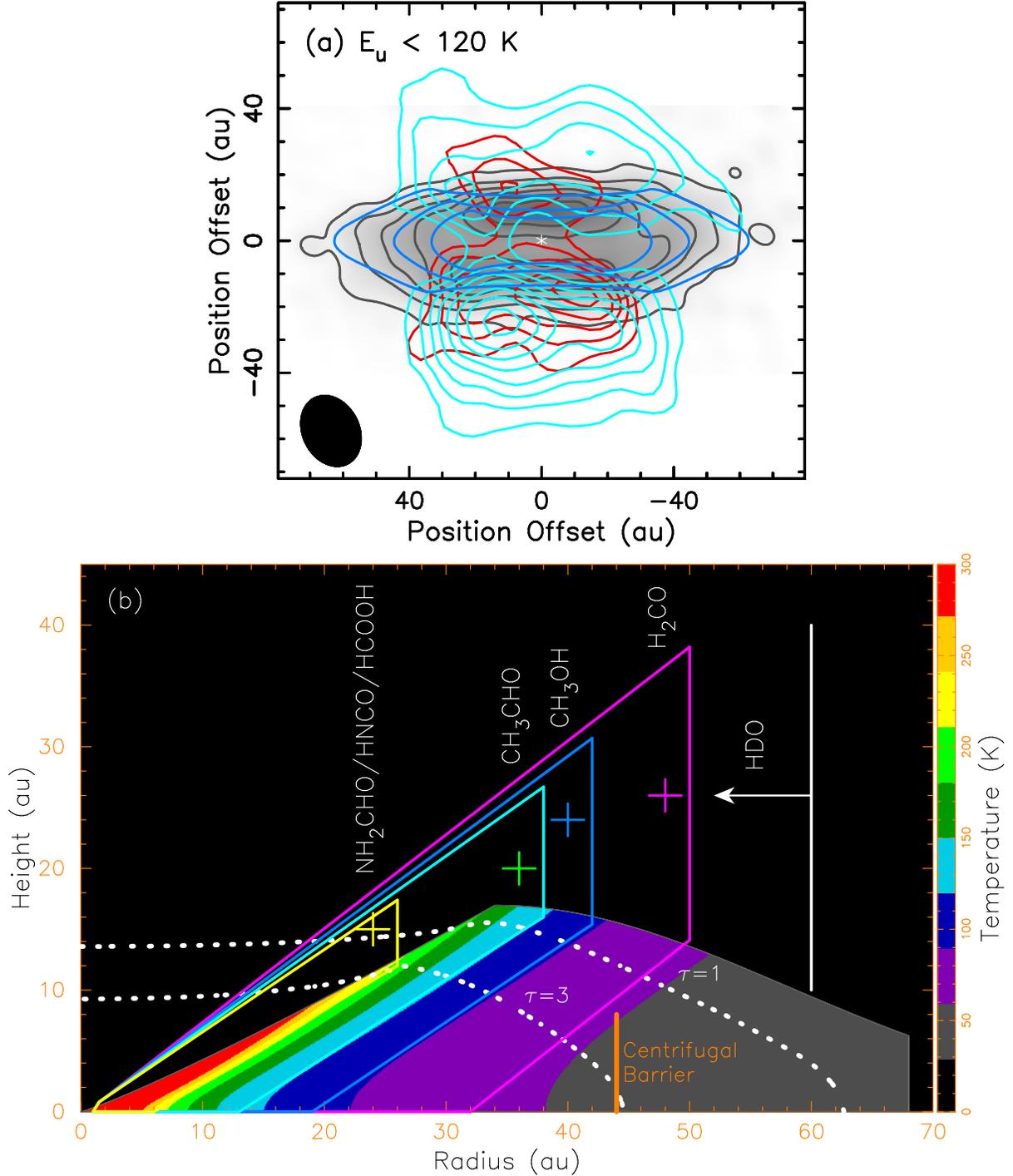
 %[!hbp]
\centering
\putfig{1}{270}{f5a.eps} %linemaps_conttau.eps}
\putfig{0.6}{270}{f5b.eps} %Poldiskmodel2.eps}
\figcaption[]
{(a) Continuum map (black contours and gray image, same as Figure
\ref{fig:contCOMs}c), continuum optical depth (blue contours, with contour levels 1, 3, and 5) at $\lambda
\sim 1.33$ mm (or 226 GHz), \NHtCHO{} (red contours, same as Figure
\ref{fig:contCOMs}a) and \CHtOH{} (cyan contours, same as Figure
\ref{fig:contCOMs}j) total line intensity maps.  (b) Stratification of
various molecules in the disk atmosphere plotted on the temperature
structure of the dusty disk adopted from \citet{Lee2021Pdisk}.  The crosses
mark the outer radius and vertical height measured from the observations.
The upper limit of the outer radius of HDO estimated at 60 au resolution by \citet{Codella2018}
is also marked.
\label{fig:conttau}}
\end{figure}

\newcounter{mtable}[section]
\newenvironment{mtable}[1][]{\refstepcounter{mtable}\par\medskip  
   \noindent \textbf{Table~\themtable. #1} \rmfamily}{\medskip}

% for row spacing in table
\renewcommand{\arraystretch}{0.6}

\begin{table}
\scriptsize
\centering
\begin{mtable}
%\caption{Observation Logs}
\bf Line Properties from Splatalogue\\
\label{tab:lines}
\end{mtable}
\begin{tabular}{llcccrc}
\hline
Transition & Frequency & log($A_{ul}$) & $E_{u}$ &$g_u$ & $W^a$    & Remarks\\
QNs        & (MHz)     & (s$^{-1}$)    & (K)     &      &(K \vkm{})&\\
\hline\hline
\NHtCHO{} 10( 1, 9)- 9( 1, 8) & 218459.21 & -3.126 &  60.812 & 21 & 119    & CDMS \\ % contaminated with other lines, e.g Ethanol and others?

\NHtCHO{} 11( 2,10)-10( 2, 9) & 232273.64 & -3.054 &  78.949 & 23 & 120    & CDMS \\

\NHtCHO{} 11( 8, 3)-10( 8, 2) & 233488.88 & -3.360 & 257.724 & 23 & 29$^m$ & CDMS \\ % overlapped with 233.49267790, with 4.87 km/s difference
\NHtCHO{} 11( 8, 4)-10( 8, 3) & 233488.88 & -3.360 & 257.724 & 23 & 29$^m$ & CDMS \\ % overlapped with 233.49267790

%\NHtCHO{} 11( 9, 2)-10( 9, 1) & 233492.67 & -3.513 & 308.237 & 23 & 17$^B$& CDMS \\ % overlapped with 233.49806500, also have 2 additional lines see JPL
%\NHtCHO{} 11( 9, 3)-10( 9, 2) & 233492.67 & -3.513 & 308.237 & 23 & 17$^B$& CDMS \\ % overlapped with 233.49806500, also have 2 additional lines see JPL

\NHtCHO{} 11( 7, 4)-10( 7, 3) & 233498.06 & -3.258 & 213.124 & 23 & 42     & CDMS \\ % overlapped with 233.49267790
\NHtCHO{} 11( 7, 5)-10( 7, 4) & 233498.06 & -3.258 & 213.124 & 23 & 42     & CDMS \\ % overlapped with 233.49267790

%\NHtCHO{} 11(10, 1)-10(10, 0) & 233505.51 & -3.793 & 364.649 & 23 & ? & CDMS \\ % overlapped with 233.49806500
%\NHtCHO{} 11(10, 2)-10(10, 1) & 233505.51 & -3.793 & 364.649 & 23 & ? & CDMS \\ % overlapped with 233.49806500

\NHtCHO{} 11( 6, 6)-10( 6, 5) & 233527.79 & -3.186 & 174.449 & 23 & 38     & CDMS \\ % 2 lines
\NHtCHO{} 11( 6, 5)-10( 6, 4) & 233527.79 & -3.186 & 174.449 & 23 & 38     & CDMS \\ % 2 lines 
\NHtCHO{} 11( 5, 7)-10( 5, 6) & 233594.50 & -3.133 & 141.714 & 23 & 56$^m$ & CDMS \\ % 2 lines
\NHtCHO{} 11( 5, 6)-10( 5, 5) & 233594.50 & -3.133 & 141.714 & 23 & 56$^m$ & CDMS \\ % 2 lines
\NHtCHO{} 11( 4, 8)-10( 4, 7) & 233734.72 & -3.093 & 114.932 & 23 & 91     & CDMS \\
\NHtCHO{} 11( 4, 7)-10( 4, 6) & 233745.61 & -3.093 & 114.933 & 23 & 132    & CDMS \\
\NHtCHO{} 11( 3, 9)-10( 3, 8) & 233896.57 & -3.064 &  94.110 & 23 & 91     & CDMS \\
\NHtCHO{} 11( 3, 8)-10( 3, 7) & 234315.49 & -3.062 &  94.158 & 23 & 96     & CDMS \\
\\
%\multicolumn{6}{c}{HNCO (with $E_u < 500$ K and $\log(A) > -4.2$ CDMS)}\\
HNCO	10( 1,10)-9( 1, 9) & 218981.00 & -3.847 & 101.078 & 21 & 212       & CDMS \\
HNCO	10( 3, 8)-9( 3, 7) & 219656.76 & -3.920 & 432.959 & 21 & 41$^m$    & CDMS \\ % 2 lines, very faint
HNCO	10( 3, 7)-9( 3, 6) & 219656.77 & -3.920 & 432.959 & 21 & 41$^m$    & CDMS \\ % 2 lines, very faint
HNCO	10( 2, 9)-9( 2, 8) & 219733.85 & -3.871 & 228.284 & 21 & 101       & CDMS \\ % overlapped with next -4.56 km/s
HNCO	10( 2, 8)-9( 2, 7) & 219737.19 & -3.871 & 228.285 & 21 & 121       & CDMS \\ % overlapped with previous
HNCO	10( 0,10)-9( 0, 9) & 219798.27 & -3.832 &  58.019 & 21 & 192$^b$   & CDMS \\ % likely optically thick
HNCO	10( 1, 9)-9( 1, 8) & 220584.75 & -3.837 & 101.502 & 21 & 198       & CDMS \\
%HNCO	28( 1,28)-29( 0,29)& 231873.25 & -4.175 & 469.868 & 57 & 103 & CDMS \\ % ? very faint, some jet-like emission?
\\
%\multicolumn{6}{c}{H2CO (with $E_u < 500$ K and $\log(A) > -4.0$ CDMS)}\\
\HtCO{}	3( 0, 3)- 2( 0, 2) & 218222.19 & -3.550 & 20.956  & 7 & 266$^b$ & CDMS \\
\HtCO{}	3( 2, 2)- 2( 2, 1) & 218475.63 & -3.803	& 68.093  & 7 & 230 & CDMS \\
\\
%\multicolumn{6}{c}{D2CO (with $E_u < 500$ K and $\log(A) > -4.0$ CDMS)}\\
\DtCO{}	4(0,4) - 3(0,3) & 231410.23 & -3.45914  & 27.88284 & 18 & 67$^b$ & CDMS \\
\DtCO{}	4(2,3) - 3(2,2) & 233650.44 & -3.57046  & 49.62595 & 18 & 104 & CDMS \\ % blended with CH3OCHO
\\
%\multicolumn{6}{c}{D2CO (with $E_u < 500$ K and $\log(A) > -4.0$ CDMS)}\\
H$_2\,^{13}$CO	3( 1, 2)- 2( 1, 1) & 219908.52 & -3.59109  & 32.93810  & 21 &120 & CDMS \\
%\\
%\multicolumn{6}{c}{D2CO (with $E_u < 500$ K and $\log(A) > -4.0$ CDMS)}\\
%DCN	J= 3 - 2 & 217238.53 & -3.33964    &    20.85164  & 21 & & CDMS \\
\\
\CHtOH{} 5( 1) - 4( 2) E1 vt=0  & 216945.52  & -4.915 &   55.871 & 11 &  347$^b$ &  JPL \\
\CHtOH{} 6( 1) - - 7( 2) - vt=1 & 217299.20  & -4.367 &  373.924 & 13 &  155 &  JPL \\
\CHtOH{} 20( 1) -20( 0) E1 vt=0 & 217886.50  & -4.471 &  508.375 & 41 &   89 &  JPL \\
\CHtOH{}  4( 2) - 3( 1) E1 vt=0 & 218440.06  & -4.329 &   45.459 &  9 &  263$^b$ &  JPL \\
\CHtOH{}  8( 0) - 7( 1) E1 vt=0 & 220078.56  & -4.599 &   96.613 & 17 &  347$^b$ &  JPL \\
\CHtOH{} 10(-5) -11(-4) E2 vt=0 & 220401.31  & -4.951 &  251.643 & 21 &  256 &  JPL \\
\CHtOH{} 10( 2) - - 9( 3) - vt=0& 231281.11  & -4.736 &  165.347 & 21 &  240$^b$ &  JPL \\
\CHtOH{} 10( 2) + - 9( 3) + vt=0& 232418.52  & -4.729 &  165.401 & 21 &  295$^b$ &  JPL \\
\CHtOH{} 18( 3) + -17( 4) + vt=0& 232783.44  & -4.664 &  446.531 & 37 &  214 &  JPL \\
\CHtOH{} 18( 3) - -17( 4) - vt=0& 233795.66  & -4.658 &  446.580 & 37 &  230 &  JPL \\
\CHtOH{}  4( 2) - - 5( 1) - vt=0& 234683.37  & -4.734 &   60.923 &  9 &  285$^b$ &  JPL \\
\CHtOH{}  5(-4) - 6(-3) E2 vt=0 & 234698.51  & -5.197 &  122.720 & 11 &  195$^b$ &  JPL \\
\\
%\CHtCHO{} 12( 4, 9)-11( 4, 8) A, vt=0 & 231456.74 & -3.409 & 108.352 &	50 & & JPL \\
%\CHtCHO{} 12( 4, 8)-11( 4, 7) A, vt=0 & 231467.50 & -3.409 & 108.353 &	50 & & JPL \\
\CHtCHO{} 12( 4, 8)-11( 4, 7) E, vt=0 & 231484.37 & -3.409 & 108.289 &	50 & 29 & JPL \\
\CHtCHO{} 12( 4, 9)-11( 4, 8) E, vt=0 & 231506.29 & -3.409 & 108.251 &	50 & 111 & JPL \\
%\CHtCHO{} 12( 3,10)-11( 3, 9) A, vt=0 & 231595.27 & -3.385 &  92.579 &	50 & & JPL \\
\CHtCHO{} 12( 3,10)-11( 3, 9) E, vt=0 & 231748.71 & -3.388 &  92.510 &	50 & 65 & JPL \\
\CHtCHO{} 12( 3, 9)-11( 3, 8) E, vt=0 & 231847.57 & -3.387 &  92.610 &	50 & 46 & JPL \\
\CHtCHO{} 12( 3, 9)-11( 3, 8) A, vt=0 & 231968.38 & -3.383 &  92.624 &	50 & 116 & JPL \\
\CHtCHO{} 12( 2,10)-11( 2, 9) E, vt=0 & 234795.45 & -3.352 &  81.864 &	50 & 54 & JPL \\
\CHtCHO{} 12( 2,10)-11( 2, 9) A, vt=0 & 234825.87 & -3.351 &  81.842 &	50 & 50 & JPL \\
\hline
\end{tabular}
\mbox{}\\
$a$: Integrated line intensities (see text for the definition) measured from the lower disk atmosphere.
Except for \CHtOH{} which used the values at the emission peak position, 
they are the mean values averaging over a rectangular region (with a size of
\arcsa{0}{17}$\times$\arcsa{0}{05} covering most of the emission) centered at
the lower atmosphere. In this column, the line intensities commented with ``m" are the mean values 
obtained by averaging over 2 or more lines with similar $E_u$ and log $A_{ul}$ for better   
measurements. $b$: likely optically thick and thus ignored in the fitting of the rotation diagram
and calculation of column density.
The line intensities here are assumed to have an uncertainty of 40\%.
\end{table}

%M: the lines used to make the total line intensity maps.
%V: the lines used to make the PV diagrams.
%R: the lines used for rotation diagrams.

%$t$:  the lines are tentatively
%detected with about 2 $\sigma$ detection.

\begin{table}
\small
\centering
\begin{mtable}
%\caption{Observation Logs}
\bf Column Densities and Abundances in the Lower Disk Atmosphere\\
\label{tab:colabun}
\end{mtable}
\begin{tabular}{llrr}
\hline
Species & Excitation Temperature & Column Density & Abundance$^\dagger$\\
& (K) & (\cms) & \\ 
\hline\hline
\CHtOH{}$^a$     &$92^{+48}_{-37}$&$1.5^{+4.5}_{-0.8}  \times 10^{18}$ cm$^{-2}$& $1.4^{+4.2}_{-0.7} \times 10^{-7}$                           \\
\NHtCHO{}$^b$    &$221\pm132$& \scnum{($5.2\pm2.6$)}{15} & \scnum{($4.8\pm2.4$)}{-10}\\
 HNCO$^b$        &$231\pm100$ & \scnum{($1.8\pm0.5$)}{16} & \scnum{($1.7\pm0.9$)}{-9} \\
\HtCO{}$^c$  &$60\pm20^e$ & $\gtrsim$\scnum{($1.4\pm0.7$)}{16} & $\gtrsim$ \scnum{($1.3\pm0.7$)}{-9} \\
\HtCO{}$^d$  &            & \scnum{($4.5\pm2.3$)}{16} & \scnum{($4.2\pm2.1$)}{-9} \\
\HttCO{}     &$60\pm20^e$ & \scnum{($9.0\pm4.5$)}{14} & \scnum{($8.3\pm4.2$)}{-11} \\
\DtCO{}      &$92\pm30^f$ & $\gtrsim$\scnum{($2.4\pm1.2$)}{15} & $\gtrsim$ \scnum{($2.2\pm1.1$)}{-10} \\
\CHtCHO{}    &$100\pm50^g$ & \scnum{($8.7\pm4.4$)}{15} & \scnum{($8.0\pm4.0$)}{-10} \\
%DCN          &$150\pm50^a$ & $\gtrsim$\scnum{($4.9\pm1.6$)}{14} & \scnum{($4.5\pm1.5$)}{-11} \\
%\NHtCHO{}$^e$ &$150\pm50^c$ & \scnum{($1.6\pm0.9$)}{15} & \scnum{($4.2\pm2.4$)}{-10}\\
%\DtCO{}$^e$ & $150\pm50^c$ & \scnum{($3.0\pm1.6$)}{15} & \scnum{($7.9\pm4.2$)}{-10} \\
\hline
\end{tabular}
\mbox{}\\
$a$: Mean temperature and column density derived from non-LTE LVG calculation and rotation diagram.\\
$b$: Temperature and column density derived from rotation diagram.\\
$c$: Column density derived from \HtCO{} line.\\
$d$: Column density derived from \HttCO{} line, assuming [$^{12}$C]/[$^{13}$C] ratio of 50.\\
$e$: Temperature assumed to be 60 K.\\
$f$: Mean temperature assumed to be the same as \CHtOH{}.\\
$g$: Temperature assumed to be 100 K.\\
$\dagger$: Abundance derived by dividing the column densities of the molecules by
the H$_2$ column density in the disk atmosphere, which is $\sim$ \scnum{1.08}{25} \cms{} (see text).\\
%$a$: Assuming an excitation temperature (rotational temperature) of $150\pm50$ K,
%which is the mean value derived from the \CHtOH{} and \CHtDOH{} lines.
\end{table}

\begin{table}
\small
\centering
\begin{mtable}
%\caption{Observation Logs}
\bf Binding Energy (BE) and Sublimation Temperature (T$_\textrm{\scriptsize sub}$)
in Amorphous Solid Water Ice\\
\label{tab:BE}
\end{mtable}
\begin{tabular}{lcccc}
\hline
Species & Binding Energy & T$_\textrm{\scriptsize sub}$ & Outer Radius & References\\
& (K) & (K) & (au) & \\ 
\hline\hline
\CHtCHO{}    &2809-6038(4423)  & 48-102(75) & 36 & Ferrero+2022\\
\HtCO{}      &3071-6194(4632)  & 52-104(78) & 48 & Ferrero+2020 \\
HNCO         &2400-8400(4800)  & 41-140(81) & 24 & Song \& Kastner 2016 \\
\CHtOH{}     &3770-8618(6194)  & 64-144(104) & 40 & Ferrero+2020  \\
HCOOH        &5382-10559(7970) & 91-176(133) & 24 & Ferrero+2020 \\
\NHtCHO{}    &5793-10960(8376) & 97-183(140) & 24 & Ferrero+2020 \\
NH$_3$       &4314-7549(5931)  & 73-126(100) & -- & Ferrero+2020 \\
H$_2$O       &3605-6111(4858)  & 61-103(82) &  60$^a$ & Ferrero+2020 \\
\hline
\end{tabular}
\mbox{}\\ \mbox{}\\
$a$: The outer radius is given by that of HDO mapped at 60 au resolution \citep{Codella2018}.\\
The numbers in the parenthesis are the mean values, except for HNCO, for
which it is a value for maximum sublimation. 
The sublimation temperature (T$_\textrm{\scriptsize sub}$) is calculated for an age of 
10$^6$ yr, appropriate for a young protoplanetary disk. 
Adopting a shorter time of 10$^5$ yr, appropriate for a 
Class 0/I protostellar system, would increase the sublimation temperature by 
a few degrees.  
Note that the computed BE values, notably those of acetaldehyde and ammonia,
may be slightly at odds with the published experimental ones, which depend
on the structure of the ices as well as the distribution of the species
population on the ices.  For this reason, we chose to stick to the BEs
computed by the same authors, Ferrero et al.,
(with the exception of HNCO because these authors did not
compute it), to make the comparison possibly more reliable.
\end{table}

% HNCO Binding Energy in Song Table 1: 48.1,27.9,80.3,52.1 kJ/mol
% 1 J= kg m^2/s^2= 10^7 erg, 1 mol = 6.022e23
% boltzamm const=1.38e-16 g s^-2 K^-1
% kJ/mol => 1e10/6.022e23/1.38e-16 K = 120.33 K
% ==3357-9662(6509)

\setcounter{mfigure}{0}
\renewcommand{\themfigure}{S\arabic{mfigure}}
\renewenvironment{mfigure}[1][]{\refstepcounter{mfigure}\par\medskip
   \noindent \textbf{Extended Data Figure~\themfigure. #1} \rmfamily}{\medskip}

\end{document}